\documentstyle[prl,aps]{revtex}
\input BoxedEPS.tex
\SetepsfEPSFSpecial
\HideDisplacementBoxes

\begin{document}

\twocolumn[\hsize\textwidth\columnwidth\hsize
          \csname @twocolumnfalse\endcsname
\title{Transport and magnetic properties in multi-walled carbon 
nanotube ropes: Evidence for superconductivity above room temperature } 
\author{Guo-meng Zhao$^{*}$} 
\address{ Department of Physics, Texas 
Center for Superconductivity and Advanced Materials, University of Houston, Houston, Texas 
77204, USA}

\maketitle
\widetext
\vspace{0.3cm}

\begin{abstract}
Detailed analyses are made on previously published data for 
multi-walled carbon nanotubes.  The field dependence of the Hall 
voltage, the temperature dependence of the Hall coefficient, and the 
magnetoresistance effect (Phys.  Rev.  Lett.  72, 697 (1994)) can all 
be consistently explained in terms of the coexistence of physically 
separated tubes and Josephson-coupled superconducting tubes with 
superconductivity above room temperature.  The observed temperature 
dependencies of the remnant magnetization, the diamagnetic 
susceptibility, and the conductance are consistent with 
superconductivity above room temperature, but are inconsistent with 
ferromagnetic contamination.  We also interpret the paramagnetic signal and unusual field dependence of the magnetization 
at 300 K (Phys.  Rev.  B 49, 15122 (1994)) as arising from the 
paramagnetic Meissner effect in a multiply connected superconducting 
network.  ~\\
~\\
\end{abstract}
\narrowtext
]

Finding room temperature superconductors is one of the most challenging 
problems in science.  It is generally accepted that low temperature 
superconductivity in simple metals arises from electron-phonon 
coupling which provides a weak effective attraction for two electrons 
to pair up.  This phonon-mediated pairing mechanism would lead to a 
superconducting transition temperature $T_{c}$ lower than 30 K 
according to McMillan's estimate \cite{Mc}.  The discovery of 
high-temperature superconductivity at about 40 K in electron-doped 
C$_{60}$ \cite{Gunnarsson} raises an interesting question of whether 
the unusually high-$T_{c}$ superconductivity in this carbon-based 
material is still mediated by phonons.  Alexandrov and Mott 
\cite{Alex} estimated that the highest $T_{c}$ within a strong 
electron-phonon coupling model is about $\omega/3$, where $\omega$ is 
the characteristic phonon frequency.  Because $\omega$ in graphite 
related materials is about 2400 K \cite{Saito}, it is possible to find 
room temperature superconductivity in graphite-related materials if 
the electron-phonon coupling could be substantially enhanced.  
Pokropivny \cite{Pok} argued that a whispering mode in nanotubes 
should be responsible for a strong enhancement of electron-phonon 
interaction, which might lead to room temperature superconductivity.  
A theoretical calculation showed that superconductivity as high as 500 
K can be reached through the pairing interaction mediated by undamped 
acoustic plasmon modes in a quasi-one-dimensional electronic system 
\cite{Lee}.  For a multi-layer electronic system such as cuprates and 
multi-walled carbon nanotubes (MWNTs), high-temperature 
superconductivity can occur due to an attraction of the carriers in 
the same conducting layer via exchange of virtual plasmons in 
neighboring layers \cite{Cui}.  Indeed, a strong coupling of electrons 
with high-energy ($\sim$ 2 eV) electronic excitations  in 
cuprates has been demonstrated  by well-designed optical experiments 
\cite{Little}.  These authors also show that the strong coupling between
electrons with high-energy electronic excitations along with strong 
electron-phonon coupling is primarily responsible for superconductivity 
above 100 K in cuprates \cite{Little}, in agreement with a recent work 
\cite{ZhaoPRB}.  For MWNTs, the dual character of the 
quasi-one-dimensional and multi-layer electronic structure could lead 
to a larger pairing interaction via electron-plasmon coupling and thus 
to superconductivity at higher temperatures.  Indeed, magnetic and 
electrical measurements on multi-walled nanotube ropes \cite{Zhao} 
have provided subtle evidence for superconductivity above 600~K.

Here we provide detailed  analyses on previously published  data for multi-walled 
nanotube ropes.  We can consistently explain 
the temperature dependencies of the Hall coefficient, the 
magnetoresistance effect, the remnant magnetization, the diamagnetic 
susceptibility, the conductance, and the field dependence 
of the Hall voltage \cite{Song} in terms of the coexistence of physically separated 
(PS) tubes  and  Josephson-coupled (JC) superconducting tubes 
with superconductivity above room temperature. We also interpret the paramagnetic 
signal and unusual field dependence of the magnetization at 300 K 
(Ref.~\cite{Heremans}) as 
arising from the paramagnetic Meissner effect in a multiply connected 
superconducting network.

We first discuss the temperature dependencies of the remnant 
magnetization $M_{r}$ and the diamagnetic susceptibility for 
our multi-walled nanotube ropes.  The experimental results 
from Ref.~\cite{Zhao} are reproduced in Fig.~1.  It is apparent that 
the temperature dependence of $M_{r}$ (Fig.1a) is similar to that of 
the diamagnetic susceptibility (Fig.1b) except for the opposite signs.  
This behavior is expected for a superconductor.  A $M_{r}$ was also 
observed by Tsebro {\em et al.} up to 300 K \cite{Tsebro}.  However, 
the observation of the $M_{r}$ alone does not give unambiguous 
evidence for superconductivity since $M_{r}$ could be caused by 
ferromagnetic impurities and/or ballistic transport.

We can now rule out the existence of ferromagnetic impurities.  If 
there were ferromagnetic impurities, the total susceptibility would 
tend to turn up below 120 K where the $M_{r}$ increases suddenly.  
This is because the paramagnetic susceptibility contributed from the 
ferromagnetic impurities would increase below 120 K.  In contrast, the 
susceptibility suddenly turns down rather than turns up below 120 K 
(Fig.~1b).  This provides strong evidence that the observed $M_{r}$ in 
our nanotubes has nothing to do with the presence of ferromagnetic 
impurities.
\begin{figure}[htb]
\input{epsf}
\epsfxsize 7cm
\centerline{\epsfbox{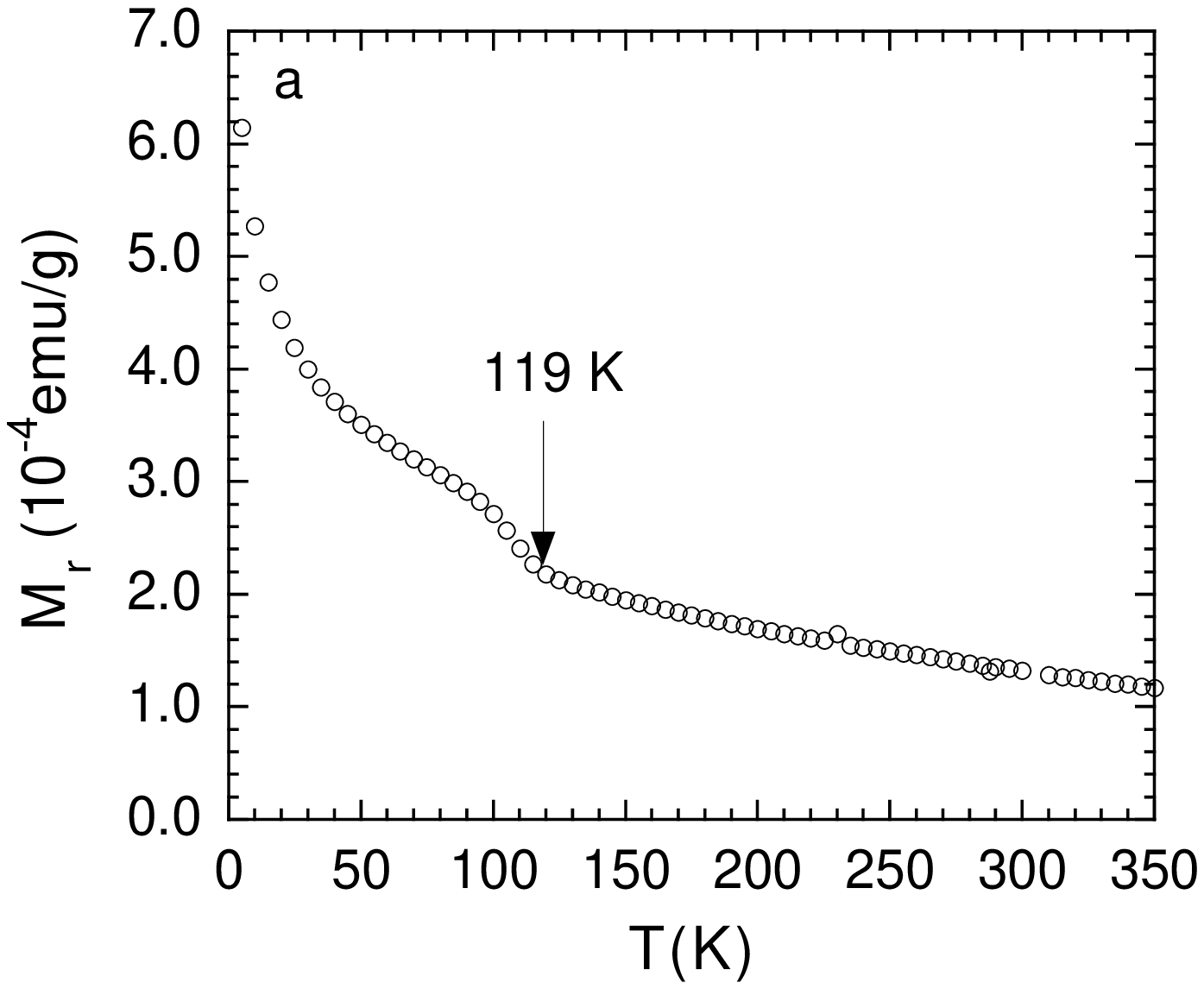}}
\vspace{0.4cm}
\input{epsf}
\epsfxsize 7cm
\centerline{\epsfbox{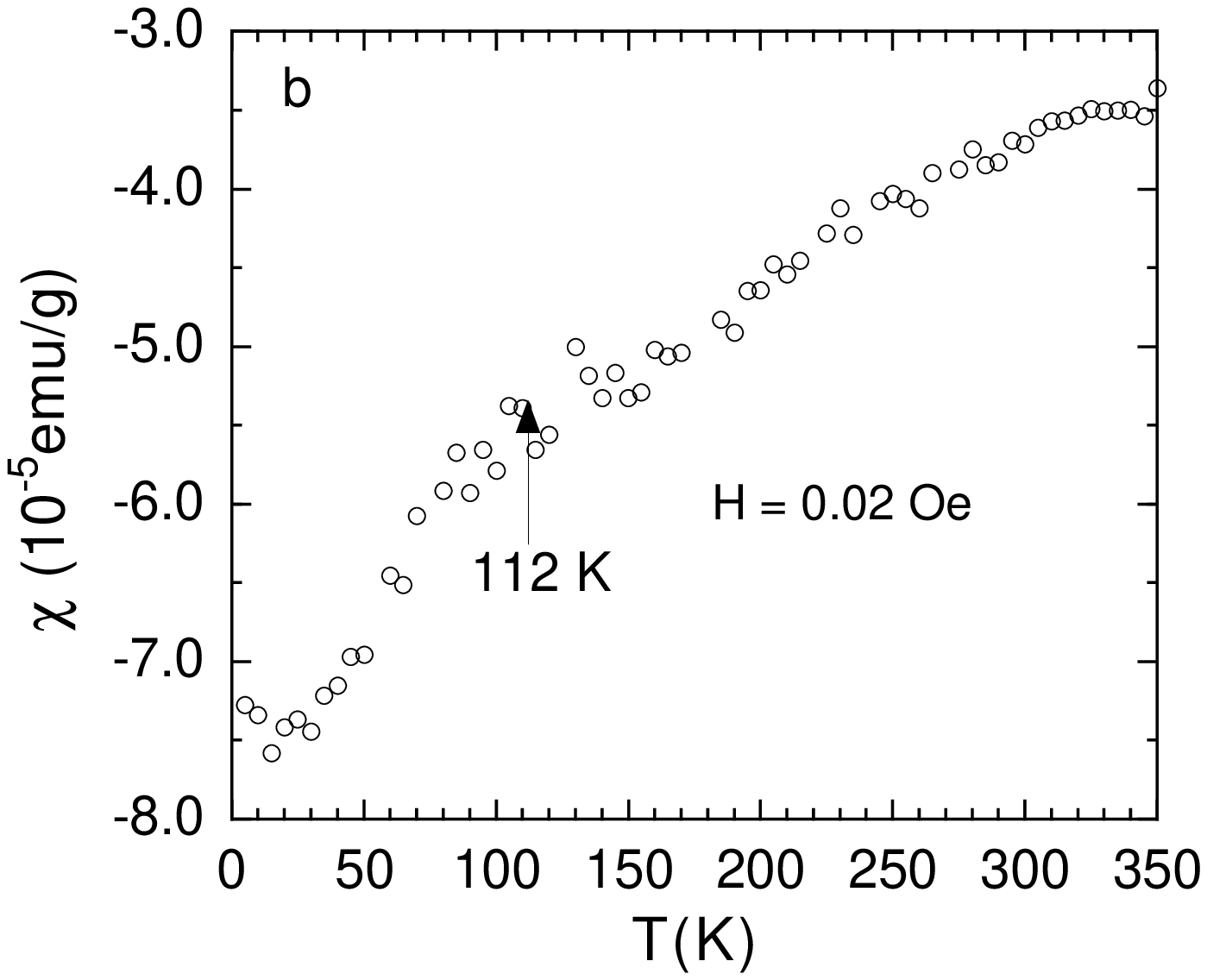}}
	\vspace{0.6cm}
	\caption [~]{a) Temperature dependence of the remnant magnetization for 
	multi-walled nanotubes.  b) The field-cooled  susceptibility as a function of 
	temperature  in a field of 0.020 Oe. After Ref.~\cite{Zhao}.}
	\protect\label{fig1}
\end{figure}
Fig.~2 shows the temperature dependence of the conductance for a multi-walled 
nanotube rope, which is reproduced from 
Ref.~\cite{Song}.  We have found the temperature dependence of the 
conductance in our nanotube ropes to be nearly the same as that shown in 
Fig.~2.  It is apparent that the slope of the conductance versus 
temperature curve for the MWNT sample starts to change below 120 K 
where both the remnant magnetization and the field-cooled diamagnetic 
signal suddenly increase.  This suggests that the magnetic properties 
are closely related to the electrical transport of the nanotubes.

Alternatively, assuming perfect conductivity, the change in the slope of the 
conductance below 120 K could be due to the increase in the ballistic 
conduction channels.  However, this scenario cannot consistently 
account for the observed increase of the field-cooled diamagnetic 
signal below 120 K since perfect conductors cannot expel magnetic flux 
in the field-cooled condition.
\begin{figure}[htb]
\input{epsf}
\epsfxsize 7cm
\centerline{\epsfbox{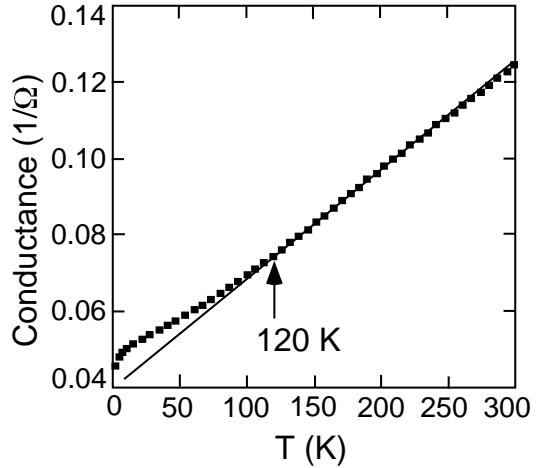}}
	\vspace{0.6cm}
	\caption [~]{Temperature dependence of the conductance for a 
	multi-walled nanotube rope, which is reproduced from 
	Ref.~\cite{Song}. We have found nearly the same temperature dependence of the 
conductance in our nanotube ropes. }
	\protect\label{fig2}
\end{figure}
In Fig.~3, we show the temperature dependence of  the Hall coefficient 
(Fig.~3a) and the field 
dependence of the Hall voltage (Fig.~3b) for a multi-walled nanotube 
rope.  These figures are again reproduced from Ref.~\cite{Song}.  It is 
striking that the Hall coefficient increases rapidly below about 120 K 
where the slopes of all three quantities in Fig.~1 and Fig.~2 
suddenly change.  The fact that such a strong temperature dependence 
below 120 K was not seen in physically separated tubes 
\cite{Baum,Chau} suggests that this is not an intrinsic property of a 
single tube, but associated with the coupling of the tubes.  Below we 
will interpret these data in a consistent way by considering the 
coexistence of physically separated (PS) tubes and Josephson-coupled 
(JC) superconducting tubes with superconductivity above room 
temperature.
 
It is well known that the carbon nanotubes have two types of electronic 
structures depending on the chirality \cite{Saito1,Ajiki}, which is 
indexed by a chiral vector $(n,m)$: $n -m = 3N +\nu$, where $N, n, m$ 
are the integers, and $\nu = 0, \pm 1$.  Tubes with $\nu$ = 0 are 
metallic while undoped tubes with $\nu$ = $\pm$1 are semiconductive.  
Multi-walled nanotubes consist of at least two concentric shells which 
could have different chiralities.  Presumably, each shell, if 
isolated and appropriately doped, should exhibit phase incoherent 
superconductivity.  If the doped shells are nested to form a MWNT such 
that there is a sufficient number of adjacent superconducting shells 
that are Josephson coupled, the single MWNT could become a phase 
coherent non-dissipative superconductor.  Similarly, if phase incoherent 
superconducting tubes are closely packed into a bundle, the bundle 
could become a phase coherent superconductor via Josephson coupling.  
It is also possible that some tubes are not superconducting due to 
insufficient doping.
\begin{figure}[htb]
\input{epsf}
\epsfxsize 7cm
\centerline{\epsfbox{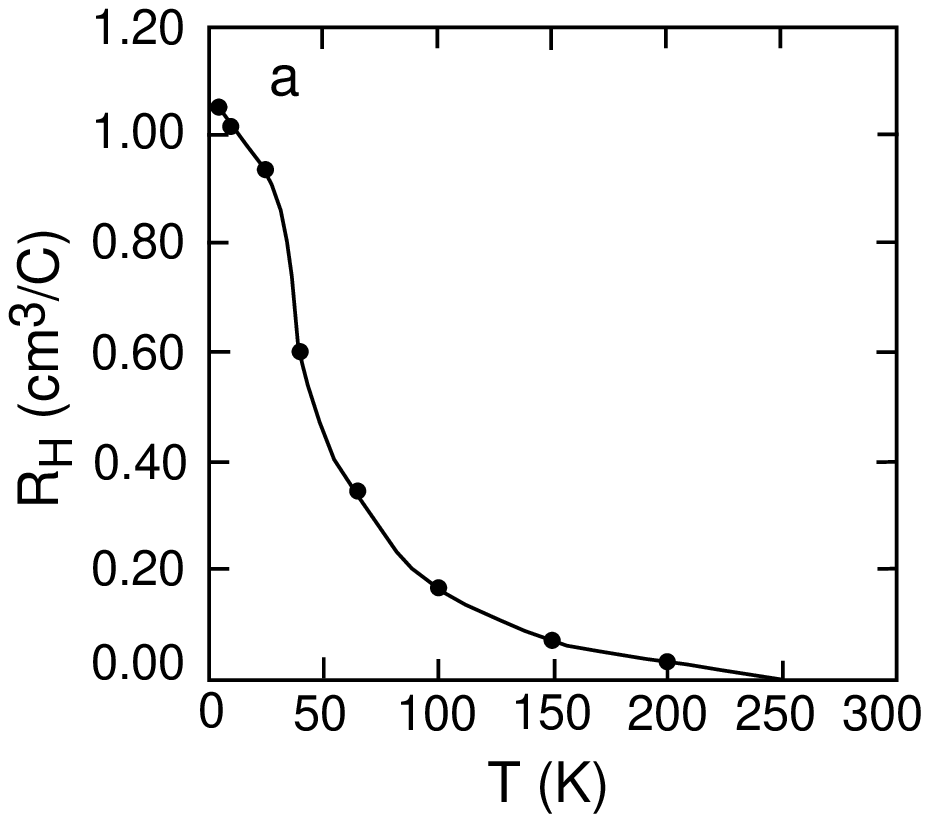}}
\vspace{0.4cm}
\input{epsf}
\epsfxsize 7cm
\centerline{\epsfbox{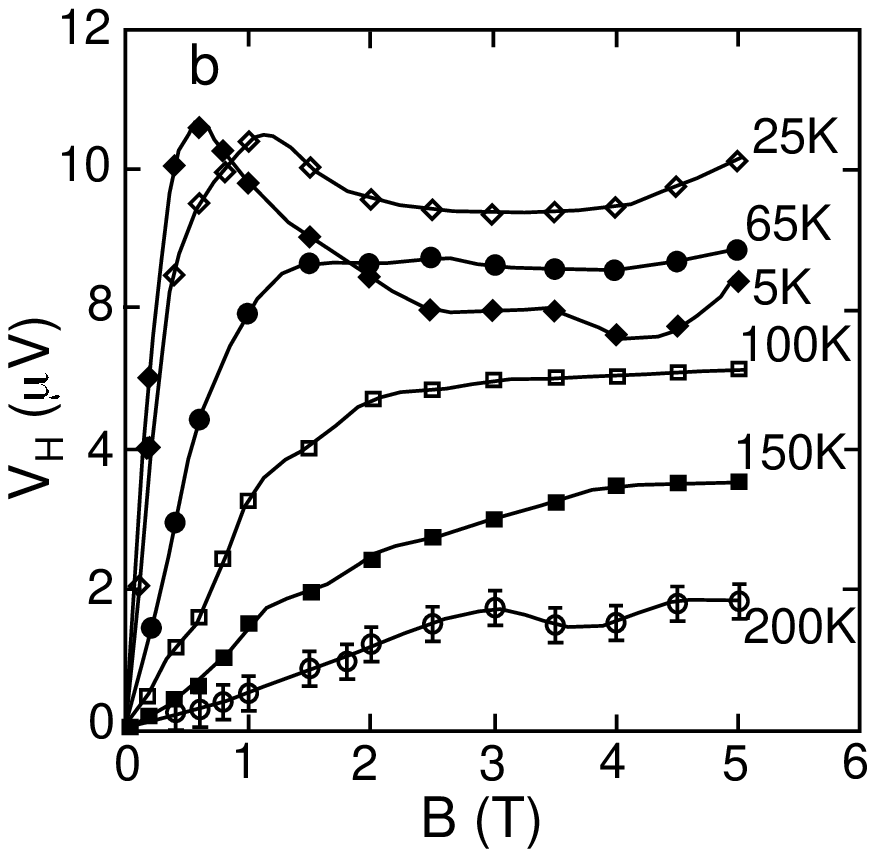}}
	\vspace{0.6cm}
	\caption [~]{a) The Hall coefficient versus temperature for a nanotube bundle.  
	b) The Hall voltage as a function of magnetic field measured at 
	different temperatures. The solid lines are drawn to guide the eye. 
	The figures are reproduced from Ref.~\cite{Song}.}
	\protect\label{fig3}
\end{figure}

We could classify the tubes in a rope into physically separated (PS) tubes  and  
Josephson-coupled (JC) superconducting tubes with superconductivity above 
room temperature.  Since the electronic properties for 
physically coupled nonsuperconducting bundles containing tubes with random 
chiralities  have no appreciable differences from those for 
physically separated nonsuperconducting tubes \cite{Maa}, we consider 
all nonsuperconducting tubes as physically separated tubes.
\begin{figure}[htb]
\input{epsf}
\epsfxsize 7cm
\centerline{\epsfbox{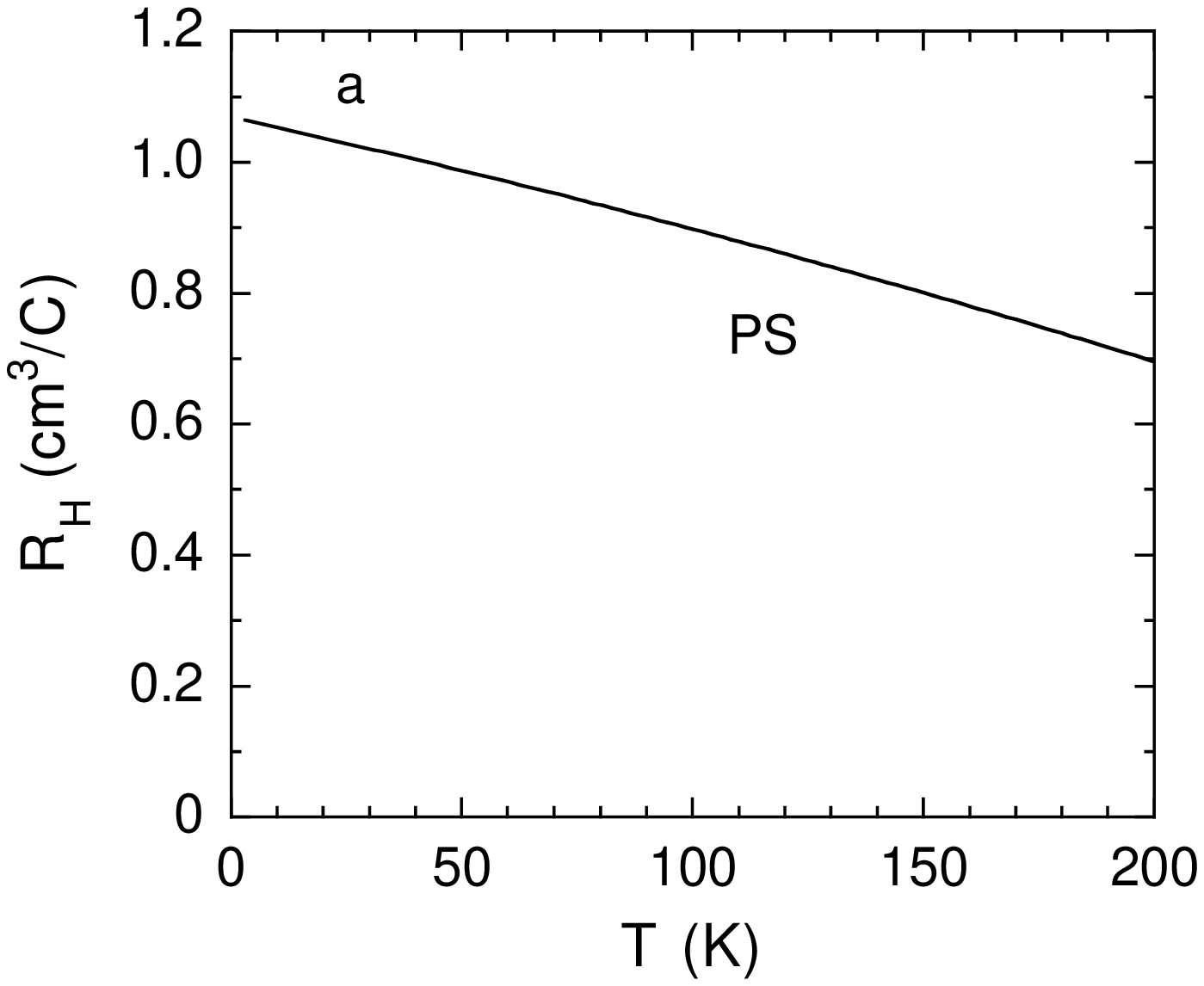}}
\input{epsf}
\epsfxsize 7cm
\centerline{\epsfbox{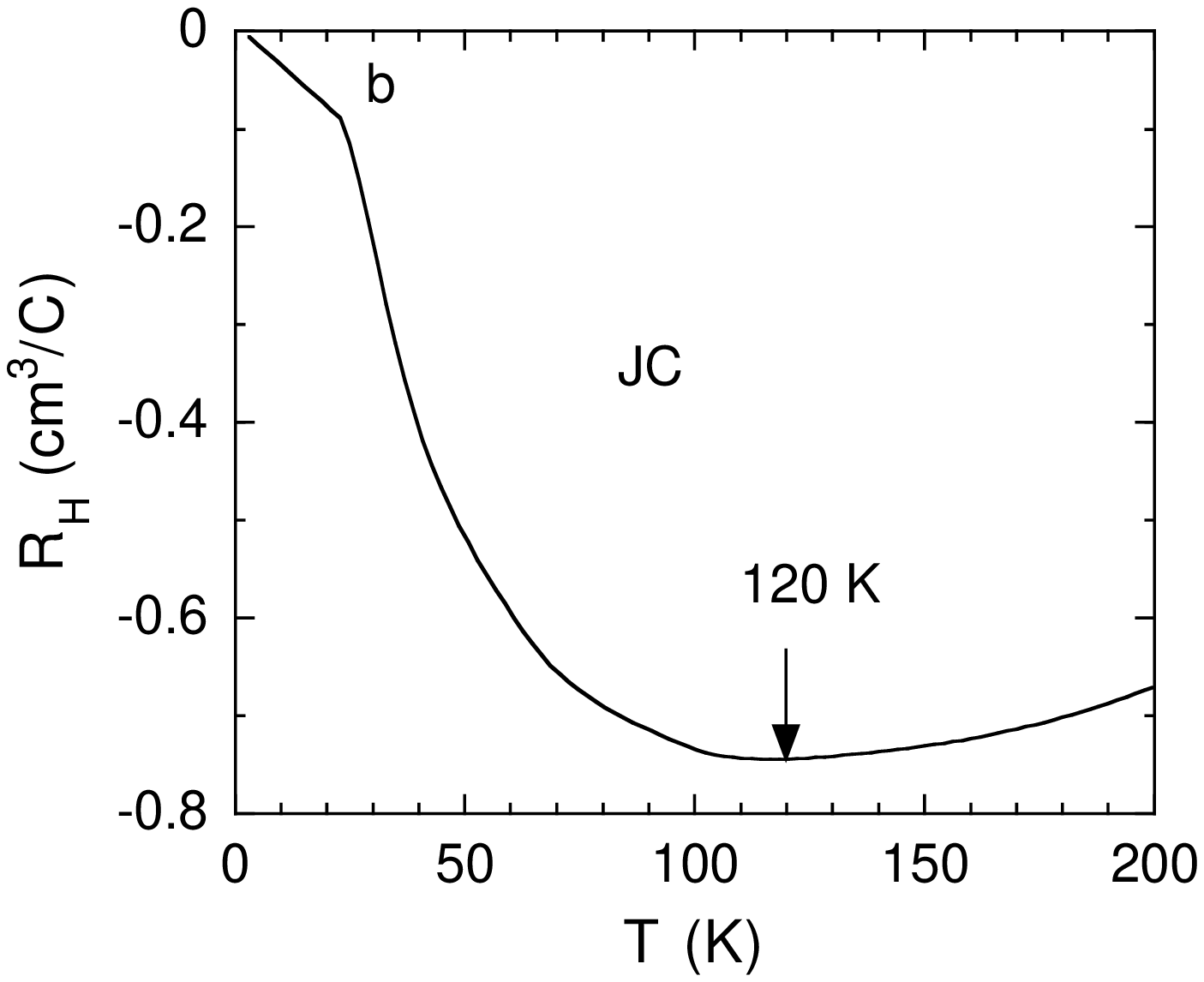}}
	\vspace{0.6cm}
	\caption [~]{a) The Hall coefficient component for physically separated  tubes 
(PS).  b) The Hall coefficient component for Josephson-coupled 
superconducting tubes (JC).  The sum of both components is taken to be 
equal to the data of Fig.~3a while the behavior shown in 
Ref.~\cite{Baum} is used to delineate the PS contribution.  }
	\protect\label{fig4}
\end{figure}
The Hall coefficient for physically separated tubes should be positive, as reported in 
Ref.~\cite{Baum}. The Hall coefficient for a single non-dissipative  
superconducting  MWNT should be zero because no vortices could be trapped 
into the single 
tube whose diameter is much smaller than the intervortex 
distance. The physically separated nonsuperconducting and phase-incoherent superconducting 
tubes should  have a positive Hall coefficient similar to that in the 
normal state. On the other 
hand, vortices can be trapped into Josephson-coupled superconducting 
tubes, leading to  a vortex-liquid state above a characteristic field 
that depends on the Josephson coupling strength.  
As seen in both cuprates and MgB$_{2}$ 
\cite{Kunchur,Jin}, the low-field Hall coefficient $R_{H}$ in the vortex-liquid state 
is negative below $T_{c}$, reaches a minimum at $T_{k}$ and then 
increases towards zero with further decreasing temperature.  Below the 
characteristic temperature $T_{k}$, vortices start to be pinned so 
that the magnitudes of the Hall conductivity, longitudinal 
conductivity, the critical current (remnant magnetization), and 
diamagnetic susceptibility increase simultaneously.  This can 
naturally explain why the diamagnetic susceptibility, the remnant 
magnetization, and the Hall coefficient simultaneously increase below 
about 120 K, as seen from Fig.~1 and Fig.~3a.

We can decompose the total Hall coefficient into two components: one for Josephson-coupled 
superconducting tubes (JC) and the other for physically separated  tubes (PS). The PS component is proportional to the measured Hall coefficient 
for physically separated  
tubes \cite{Baum} with the constraint that, at zero temperature, the magnitude of the PS 
component is equal to the total Hall coefficient. The JC component  is 
obtained by subtracting the PS component from the total 
Hall coefficient. Fig.~4 shows both the PS 
and  JC components. It is apparent that the JC 
component has a local minimum at $T_{k} \simeq$ 120 K, where the total 
Hall coefficient starts to increase rapidly (see Fig.~3a). The negative value 
of the JC component remains up to 200 K, suggesting that the 
superconducting transition temperature is far above 200 K.

\begin{figure}[htb]
\input{epsf}
\epsfxsize 7cm
\centerline{\epsfbox{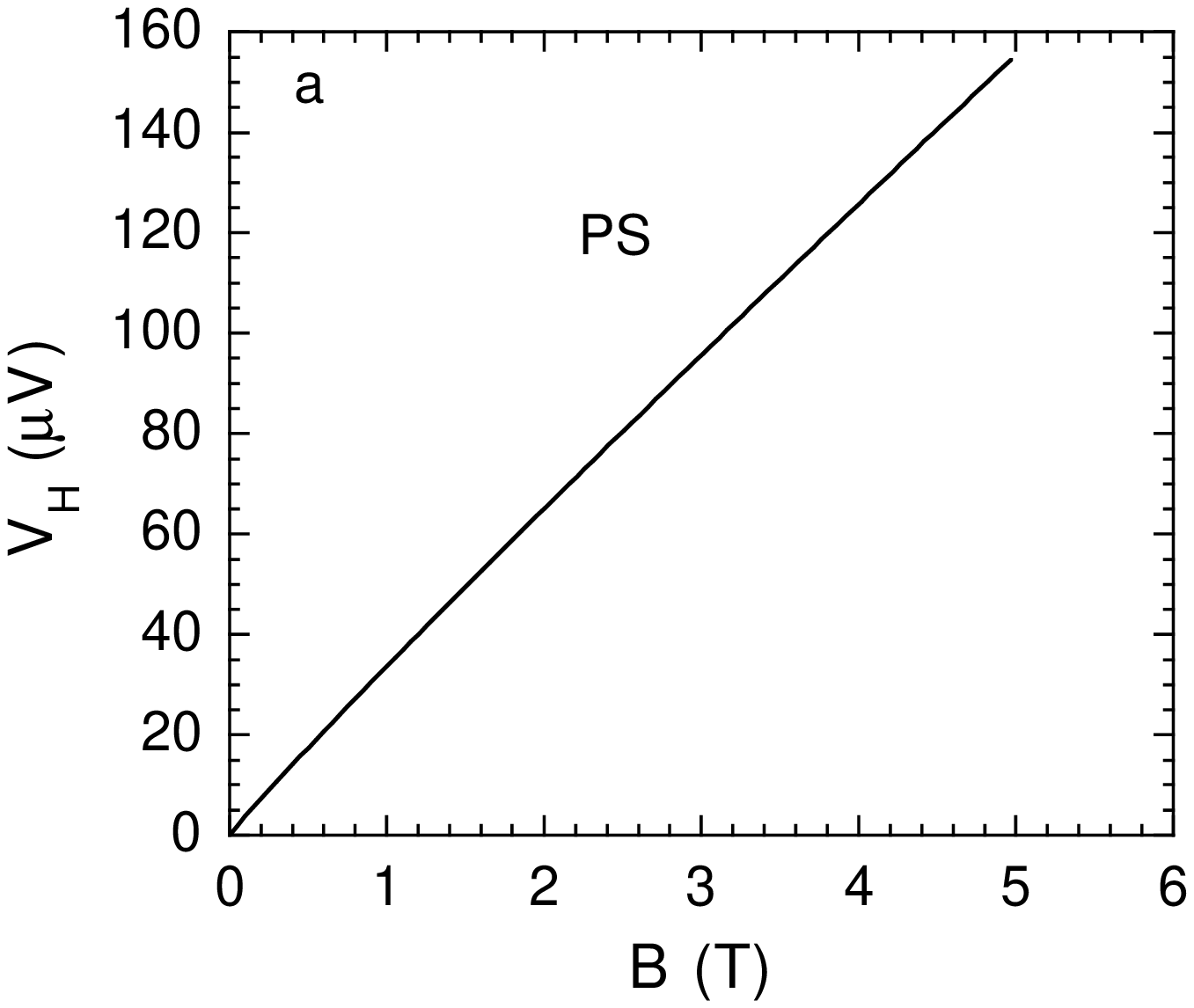}}
\vspace{0.4cm}
\input{epsf}
\epsfxsize 7cm
\centerline{\epsfbox{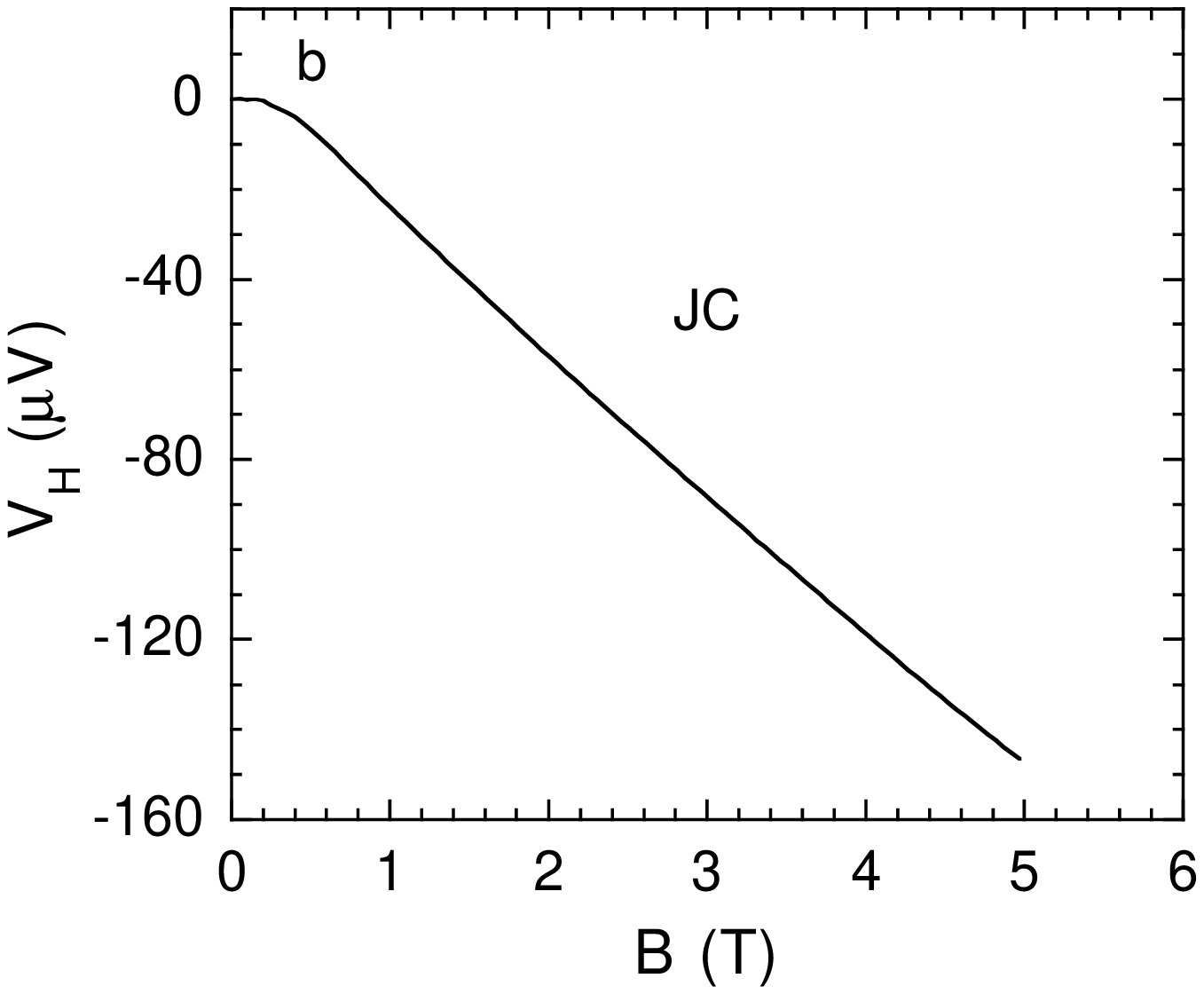}}
	\vspace{0.6cm}
	\caption {a) The Hall voltage component for physically separated  tubes 
	(PS) at 5 K.  b) The Hall voltage component for Josephson-coupled 
	superconducting tubes (JC) at 5 K. }
	\protect\label{fig5}
\end{figure}

Similarly, as seen in cuprates and MgB$_{2}$ \cite{Kunchur,Jin,Chien}, the Hall 
voltage  $V_{H}$ in the vortex-liquid state is negative, passing 
through a minimum at $B_{k}$, and then increasing towards the 
normal-state value with further increasing temperature. Below $B_{o}$, $V_{H}$ tends to zero. Interestingly, both $B_{o}$ and 
$B_{k}$ can be independently obtained from the field dependence of 
the longitudinal resistivity, as described in Ref.~\cite{Chien}.

We can also decompose the total Hall voltage into two components: one 
for Josephson-coupled superconducting tubes (JC) and the other for physically 
separated tubes (PS).  Plotted in Fig.~5a is the PS component at 5 K, 
which is proportional to the measured Hall voltage for physically 
separated tubes \cite{Baum} and matches with the low field data shown 
in Fig.~3b.  Fig.~5b shows the JC component at 5 K, which is obtained 
by subtracting the PS component from the total Hall voltage.  The 
decomposition was performed after the data in Fig.~3b were smoothed.  
We can see that the field dependence of the JC component is quite 
similar to that for cuprates and MgB$_{2}$ \cite{Kunchur,Jin,Chien} 
except that $B_{o}$ in the MWNTs is rather small, which may be due to 
a weak pinning potential.  Fig.~5b also indicates that the magnitude 
of $B_{k}$ is larger than 5 T.  As discussed below, this is in 
agreement with the longitudinal magnetoresistance data.

 Since the contribution from the physically separated tubes is 
 negligible \cite{Baum}, the longitudinal magnetoresistance at 300 K 
 mainly arises from the Josephson-coupled superconducting tubes.  From 
 the magnetoresistance data at 300 K \cite{Song} and the criterion for 
 determining $B_{k}$ \cite{Chien}, we find that $B_{k} \simeq$ 3.0 T at 
 300 K.  Using $B_{k}(T) \propto (1-T/T_{c})^{1.5}$ 
 (Ref.~\cite{Chien}), $B_{k}(5 K)$ $>$ 5 T, and $B_{k}(300 K)$ = 3.0 T, 
 we find that 300 K$<$$T_{c}$ $<$ 1070 K.

Within this two component 
model, we can  also explain the unusual 
magnetoresistance (MR) effect below 150 K.  Because physically separated 
tubes produce a 
negative MR effect at low temperatures while the Josephson-coupled 
superconducting tubes generate a 
positive MR effect, these opposing contributions from the two components can lead to a local 
minimum at certain magnetic field. This is 
indeed the case  (see Fig.~1 of 
Ref.~\cite{Song}).  At high 
temperatures, the negative MR effect contributed from the 
physically separated tubes should become weak so that the positive MR 
effect contributed from the Josephson-coupled 
superconducting tubes dominates, in agreement with the experimental results 
\cite{Zhao,Song}.

There are more experimental results that support the thesis of room 
temperature superconductivity in multi-walled nanotubes.  Fig.~6a shows the temperature dependence of the 
susceptibility for a MWNT nanotube rope in a field of  $H$ = 400 
Oe,  which is reproduced from Ref.~\cite{Heremans}. It is striking 
that the temperature dependence of the susceptibility for the MWNT rope 
is similar to that for a ceramic cuprate superconductor 
Bi$_{2}$Sr$_{2}$CaCu$_{2}$O$_{8+y}$ (BSCCO) in a field of $H$ = 0.02 
Oe (see Fig.~6b which is reproduced from Ref.~\cite{Bra}).  By 
analogy, the observation of a paramagnetic signal at 300 K could be 
explained as arising from the paramagnetic Meissner effect below the 
superconducting transition temperature, as observed in ceramic cuprate 
superconductors \cite{Bra} and in multijunction loops of conventional 
superconductors \cite{Ara}.

\begin{figure}[htb]
\input{epsf}
\epsfxsize 6cm
\centerline{\epsfbox{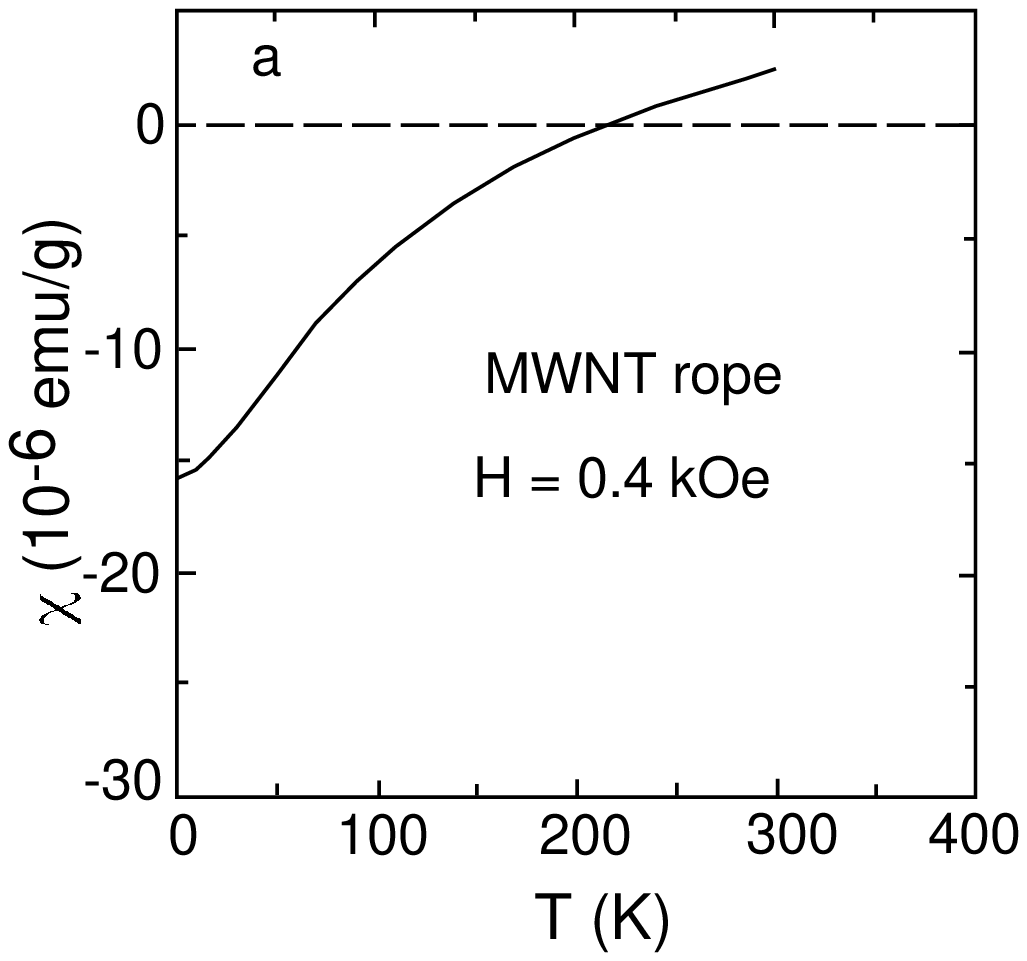}}
\vspace{0.4cm}
\input{epsf}
\epsfxsize 6cm
\centerline{\epsfbox{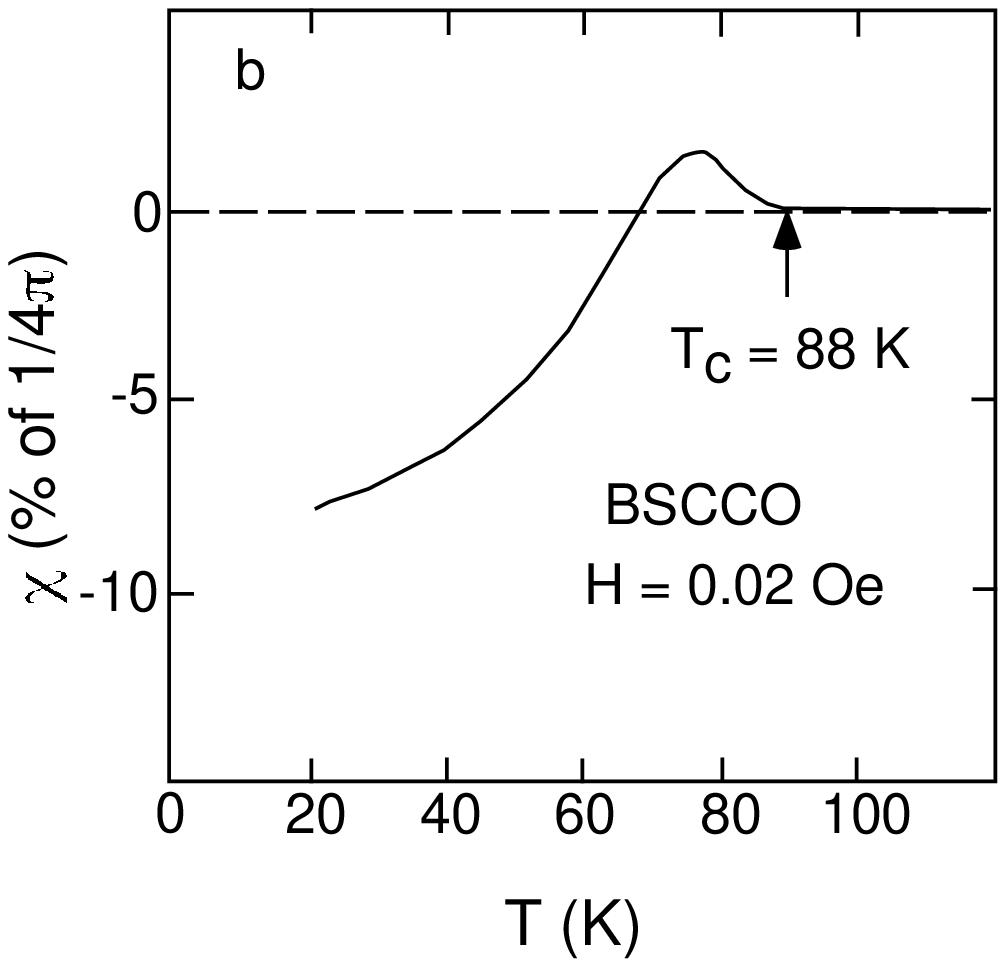}}
	\vspace{0.6cm}
	\caption [~]{a) Temperature dependence of the 
susceptibility for a MWNT nanotube rope in a field of  $H$ = 400 
Oe. The figure is reproduced from Ref.~\cite{Heremans}.  b) Temperature dependence of the 
susceptibility of  Bi$_{2}$Sr$_{2}$CaCu$_{2}$O$_{8+y}$ 
(BSCCO) in a field of $H$ = 0.02 Oe . The figure  is reproduced 
from Ref.~\cite{Bra}.  }
	\protect\label{fig6}
\end{figure}
On the other hand, the observed paramagnetic signal at 300 K and the $M(H)$ curve below $H$ = 10 kOe (see Fig.~7 of 
Ref.~\cite{Heremans}) could be compatible with the presence of ferromagnetic 
impurities.  However, such ferromagnetic impurities should be detectable 
in the high-field magnetization curve by a non-zero intercept in the 
extrapolation for $H \rightarrow$ 0. The intercept was found to be 
nearly zero at 300 K in the samples of Ref.~\cite{Heremans}.  In our 
samples \cite{Zhao} prepared from graphite rods with the same purity 
(99.9995$\%$) as the ones in Ref.~\cite{Heremans}, the intercepts are
negligible throughout the temperature range of 250 K to 400 K. For  the less 
pure C$_{60}$ and 
graphite samples of Ref.~\cite{Heremans}, the contamination of ferromagnetic 
impurities is clearly seen from the $M(H)$ curve in the high-field 
range \cite{Heremans}.  The clear ``ferromagnetic'' signal observed 
only in the low field range \cite{Heremans} is similar to that in the case of  
granular superconductors \cite{Bra}.  The ``ferromagnetic'' signal may 
be caused by a ``ferromagnetic'' ordering of elementary long-thin 
current loops \cite{Eagles}.  The critical magnetic field below which 
the ``ferromagnetic'' state is stable should depend on the number of 
filaments per current loop.  The large critical field of about 10 kOe 
in the samples of Ref.~\cite{Heremans} suggests that a current loop 
may correspond to a bundle consisting of a large number of tubes.

In summary, we have made detailed analyses on  previously published 
data for multi-walled nanotube ropes.  The observed field dependence 
of the Hall voltage, the temperature dependence of the Hall 
coefficient, and the magnetoresistance effect can be consistently 
explained in terms of the coexistence of physically separated tubes 
and Josephson-coupled superconducting tubes with superconductivity 
above room temperature.  The temperature dependencies of the remnant 
magnetization, the diamagnetic susceptibility, and the conductance are 
consistent with superconductivity rather than with ballistic transport 
or inclusion of ferromagnetic impurities.  We also interpret the 
observed paramagnetic signal and unusual field dependence of the 
magnetization at 300 K as arising from the paramagnetic Meissner 
effect in multiply connected room-temperature superconducting network.  
~\\
 ~\\
 ~\\
{\bf Acknowledgment:}  The author thanks Dr. Pieder Beeli for his critical reading and 
comments on the manuscript. The author acknowledges financial support from the State of Texas 
through the Texas Center for Superconductivity and Advanced Materials at the 
University of Houston where some of the work was completed.  ~\\
 ~\\  
* Correspondence should be addressed to gmzhao@uh.edu.

\end{document}